\documentclass[times]{ettauth}
\usepackage{tikz}

\usepackage{amsfonts}
\usepackage{amssymb}
\usepackage{latexsym}
\usepackage{cite}
\usepackage[tight,footnotesize]{subfigure}
\usepackage{color}
\usepackage{mathtools}
\usepackage[autostyle]{csquotes}
\usepackage{textcomp}
\usepackage{amsmath}
\usepackage{indentfirst}
\usepackage{moreverb}
\usepackage{setspace}
\usepackage[colorlinks,bookmarksopen,bookmarksnumbered,citecolor=red,urlcolor=red]{hyperref}
\usepackage[]{siunitx}

\usepackage{color}
\definecolor{mgreen}{rgb}{0,0.2,0.2}
\definecolor{mblue}{rgb}{0,0,0.8}
\definecolor{mred}{rgb}{0.8,0,0}

\def\volumeyear{2016}

\newcommand{\rrn}{r_{r}}
\newcommand{\renz}{r_{\text{enz}}}
\newcommand{\optr}{r_{\text{enz}}^{*}}
\newcommand{\Renz}{R_{\text{enz}}}
\newcommand{\onemicro}{\SI{1}{\micro\meter}}

\newcommand{\ts}{t_s}
\newcommand{\tend}{t_{\text{end}}}
\newcommand{\dist}{\textit{d}}
\newcommand{\hl}{\Lambda_{1/2}}

\newcommand{\hle}{\Lambda^{\SI{1}{\micro\meter}}_{1/2}}
\newcommand{\vlp}{V_{\text{lp}}}
\newcommand{\vtot}{V_{\text{totenz}}}

\newcommand{\etal}{\textit{et al.}}
\newcommand{\ap}{\textquotesingle\;\!}

\begin{document}

\runningheads{Y. J. Cho, H. B. Yilmaz, Weisi Guo and C.-B. Chae}{}

\articletype{RESEARCH ARTICLE}

\title{Effective inter-symbol interference mitigation with a limited amount of enzymes in molecular communications }

\author{Yae Jee~Cho\affil{1}, H.~Birkan~Yilmaz\affil{1}, Weisi Guo\affil{2}, and Chan-Byoung~Chae\affil{1}\corrauth}

\address{\affilnum{1}School of Integrated Technology, Yonsei Institute of Convergence Technology, Yonsei University, Korea \\ \affilnum{2}School of Engineering, University of Warwick, UK}
\corraddr{School of Integrated Technology, Yonsei Institute of Convergence Technology, Yonsei University, Korea.\\ 
E-mail: cbchae@yonsei.ac.kr}

\begin{abstract}
In molecular communication via diffusion (MCvD), the inter-symbol interference (ISI) is a well known severe problem that deteriorates both data rates and link reliability. ISI mainly occurs due to the slow and highly random propagation of the messenger molecules, which causes the emitted molecules from the previous symbols to interfere with molecules from the current symbol. An effective way to mitigate the ISI is using enzymes to degrade undesired molecules. Prior work on ISI mitigation by enzymes has assumed an infinite amount of enzymes randomly distributed around the molecular channel. Taking a different approach, this paper assumes an MCvD channel with a limited amount of enzymes. The main question this paper addresses is how to deploy these enzymes in an effective structure so that ISI mitigation is maximized. To find an effective MCvD channel environment, this study considers optimization of the shape of the transmitter node, the deployment location and structure, the size of the enzyme deployed area, and the half-lives of the enzymes. It also analyzes the dependence of the optimum size of the enzyme area on the distance and half-life.\\
\end{abstract}
\maketitle
\begin{figure*}[t]
\centering{\includegraphics[width=0.99\textwidth,keepaspectratio]
{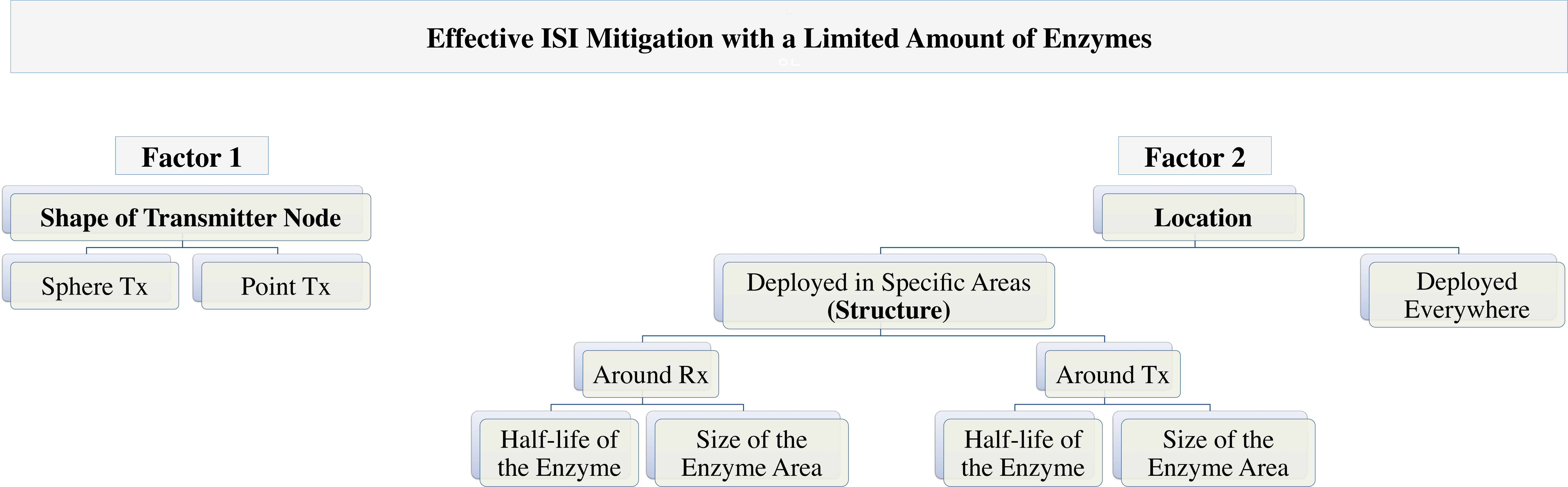}}
\caption{Steps of analyzing effective ways to deploy a limited amount of enzymes in MCvD for ISI mitigation.}  \label{fig:IntroDiag}
\end{figure*}

\section{Introduction}
As nanotechnology continually expands its\ap\,\! \!field and gains significance, researchers have produced diverse developments in nano-scale devices advance, advancing such areas such as bioscience, environmental engineering, and others. Along with such developments, researchers have studied molecular communication via diffusion (MCvD) as a probable means for nano-scale communication~\cite{akyildiz2011nanonetworksAN,nakano2013molecularC, Farsad2016ComSurv}. As for the dominantly used radio-frequency (RF) communication, nano-range is difficult to implement due to the severe path-loss~\cite{Guo2015MCVSRF}. While MCvD has better path-loss properties than RF in short-ranged communication, the high level of randomness in signal propagation creates problematic non-linear noise in macro-scale applications~\cite{KimNa2014ChaNoise}. Moreover, the heavy tail nature of the received signal causes inter-symbol interference (ISI), which is detrimental to the capacity of a MCvD channel since ISI can increase the error-rate or decrease the data-rate. Depending on the symbol duration, ISI is one or more symbols from previous symbol periods interfering with the current symbol and causing noise at the receiver node~\cite{tepekule2015isiMT,Birk2014SimStudy,kuran2012interferenceEO,kilinc2013receiverDF}.

MCvD utilizes messenger molecules as the transmitting signal between two nodes, Tx and Rx, for communication. This is the general concept, and details of the system can be diversified by characteristics such as the shape of the Tx and Rx or distance between the nodes. For a molecular-concentration based MCvD system~\cite{kuran2011modulationTF,nakano2013molecularC,KimNa2013MCvD}, the Tx either emits molecules or does nothing at each pre-decided symbol period, according to the intended message. Hence in analyzing the system capacity what is very significant is the shape of the received signal at the Rx is very significant. Since ISI occurs when the signal intended for the previous symbol does not propagate fast enough directly to the receiver. Hence, one solution is to increase the symbol duration so that the system can wait until all of the messenger molecules reach the Rx within its\ap \,\!\! symbol period. Doing so, however, decreases the data-rate. Rather than simply increasing the symbol duration, a possible method may be symbol interval optimization~\cite{KimNa2014SymInt}. Another approach is to utilize decision feedback mechanisms in the amplitude modulation method~\cite{lin2012signal,Birk2014SimStudy}. A more reasonable solution, however, and one that does not elongate the symbol period, is that of using enzymes to destroy the ISI molecules. Although this decreases signal power because enzymes also decompose the molecules that make up the current signal, the loss of power can be compensated for by lowering the decoding threshold.

Several studies have proposed different ideas to implement enzymes for ISI mitigation~\cite{kuran2013tunnel, Noel2014MCwE, Akif2015MCwE}. In ~\cite{kuran2013tunnel}, Kuran \etal \, proposed using \enquote{destroyer molecules}, similar to enzymes, to decrease the mean and variance of the hitting time distribution. Here the researchers deployed an unlimited amount of destroyer molecules inside a cylindrical tunnel structure--- a direct and restricted path between the point Tx and the sphere Rx. In~\cite{Noel2014MCwE}, Noel \etal \, also proposed using enzymes to mitigate ISI in a 3-dimensional (3D) MCvD channel with a non-absorbing receiver. An infinite quantity of enzymes were assumed to be spread throughout the channel by an infinite amount. A favourable performance in ISI mitigation was evidenced in a decreased bit-error-rate. In~\cite{Akif2015MCwE}, Heren \etal \, presented an analytical function for the hitting probability of an MCvD channel with an infinite amount of enzymes deployed everywhere. All these studies have demonstrated different approaches to using an infinite amount of enzymes for ISI mitigation. From a resource perspective, enzymes could be used more efficiently in a limited amounts. Indeed, it may not be practical to assume a deployment of an infinite amount of enzymes may not be practical.

This paper presents an analysis of effectively using a limited amount of enzymes in different system structures is presented. If an unlimited number of enzymes are available, then there is no question of \textit{where} to deploy them, as optimal ISI mitigation would result from deploying them everywhere within an appropriate concentration. In a limited enzymes situation, however, a critical factor would be to deploy them in an effective location and structure. After verifying that using enzymes produces a lower ISI than using no enzymes at all, this study compares different shapes of Tx (sphere and point). It then considers the deployment location--enzymes randomly deployed \enquote{everywhere}\footnote{Note that this scheme of deploying a limited amount of enzymes~\enquote{everywhere} is not exactly the same with the case of deploying an infinite amount of enzymes. In the case of the limited enzymes scenario, having enzymes everywhere yields to zero concentration of enzymes asymptotically. Therefore, instead of exactly deploying the enzymes everywhere we consider a sphere with a big radius for the enzyme deployment area for making it comparable with the other limited enzyme cases.} versus specific locations. Afterwards, to find the optimum case for ISI mitigation, the study compares the results from specific areas (i.e. structures) \enquote{around Rx} and \enquote{around Tx}. Lastly, we consider the specific system parameters, the size of the enzyme area and the half-lives of the enzymes, are taken into account to see which scenario most mitigates ISI. Figure~\ref{fig:IntroDiag} summarizes the main aspects of the limited enzyme deployment issue.

The paper is organized as follows. Section II gives both quantitative and qualitative descriptions of the MCvD channel and enzyme dynamics. Section III expands the MCvD concept to the limited enzymatic MCvD channel specific to our paper including topology, geometry, and scenarios of how the limited enzymes are implemented. Section IV elaborates on the simulation system used in this paper and Section V gives a specific analysis of the results. Section VI concludes the paper.

\section{System Modeling}
\subsection{Molecular Communication via Diffusion}
\begin{figure}[t]
\centering{\includegraphics[width=1\columnwidth,keepaspectratio]
{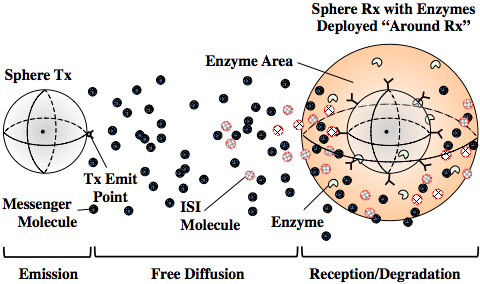}}
\caption{A MCvD system with a limited amount of enzymes deployed around Rx for a sphere Rx and Tx.}  \label{fig:MCvD}
\end{figure}

In a general MCvD system, the transmitter node emits messenger molecules which freely diffuse by Brownian motion~\cite{Nakano2012Brown, tyrrell1984diffusion, berg1993random, redner2001guide} towards the receiver. Once the molecular signal is received by the receiver it is decoded accordingly by the system's modulation scheme. Modulation can be done in different ways depending on properties such as concentration, type, and time of release of the messenger molecules~\cite{kuran2011modulationTF, KimNa2013MCvD}. Since path-loss for MCvD is proportional to ${\dist^{-3}}$ which is lower than that of RF which is ${\dist^{-2}}$, molecular communication has lower path-loss distortion when used in nano-environments~\cite{Yilmaz20143Drx,Guo2015MCVSRF}. The problem of MCvD is due to the long propagation time proportional to $\dist^2$, which is square to that of RF~\cite{Guo2015MCVSRF}. This means that molecules diffuse so slowly that they exceed their symbol period and interfere with the next symbol period\ap\!s molecules, creating ISI. Figure~\ref{fig:MCvD} shows a diagram of the MCvD system with a limited amount of enzymes deployed around the Rx.

In analyzing the MCvD system, the important factor is how the molecular signal is perceived at the receiver~\cite{Birk2014Arrivmodel}. The peak, tail, and duration of the received signal are directly related to the system\ap \!s decoding scheme, error-rate, and data-rate. For a 3D MCvD system with a point Tx and an absorbing sphere Rx, the hitting probability of sent molecules to the receiver is
\begin{align}
h(t) = \frac{\rrn}{\dist+\rrn}\frac{\dist}{\sqrt{4\pi Dt^3}}e^{-\frac{\dist^2}{4Dt}}. 
\label{hitp}
\end{align}
where $\rrn$, $\dist$, and $D$ is the receiver radius, the shortest distance between the Rx and Tx, and the diffusion coefficient, respectively~\cite{Yilmaz20143Drx}. 
The equation gives a general understanding of how molecules behave inside the channel without enzymes.
\subsection{Enzyme Dynamics}
Enzymes, in nature, are substances that catalyse and speed up reactions so that mechanisms can function properly. Catalysis is done by decomposing certain substrates to different molecules. Most of the enzymes do not just act on any substrates, but depending on the type of the enzyme, they may decompose with specificity targeting only particular types of molecules or chemical bonds. The chemical reaction is defined by:
\begin{align}
	E+S  \xrightleftharpoons[k_{-\!1}]{\,k_1\,} ES \xrightarrow{k_p} E+P
\label{enzm_ch}
\end{align}
where $E, \, S, \, ES, \, P$, and $k_n$ is the enzyme, substrate, enzyme-substrate compound, product, and rate of reactions, respectively.  By applying the law of mass action, the law which shows that the rate of reaction is proportional to the concentration of reactants~\cite{Murray2002enzchem}, to \eqref{enzm_ch}, we get the following differential equations that define the enzymatic reactions: 
\begin{align}
\begin{split}
	\frac{d[S]}{dt} 		&= -k_1[E][S]+k_{-1}[ES] \\
	\frac{d[E]}{dt} 		&= -k_1[E][S]+k_{-1}[ES]+k_p[ES] \\
	\frac{d[ES]}{dt} 		&= k_1[E][S]-k_{-1}[ES]-k_p[ES]
\label{enzm_1}
\end{split}  
\end{align}
where $[\cdot]$ corresponds to the concentration operator. In this paper, a specific case of enzymatic reaction is considered under the following assumptions:
\begin{itemize}\itemsep2pt
  \item $k_p  \xrightarrow \, \infty$ and $k_{-1} \xrightarrow \, 0$, $\quad \therefore S \xrightarrow \, P$
  \item $[ES] \xrightarrow \, 0$, $\quad \therefore [ES] = d[ES]/dt = 0$
\end{itemize}
These assumptions imply a very fast enzymatic reaction, which can be realized by selecting the appropriate pairs of enzymes and messenger molecules. Applying the assumptions to \eqref{enzm_1}, we get
\begin{align}
\begin{split}
	\frac{d[S]}{dt}     	&= -k_1[S][E]  \\
	\frac{d[ES]}{dt} &= \frac{d[E]}{dt}     	= 0. 
	\label{enzm_2}
\end{split}  
\end{align}
By solving \eqref{enzm_2}, the concentration of messenger molecules (substrate) at time $t$, namely $C(t)$, with the initial substrate concentration $C_0$, is derived as an exponential decay function,
\begin{align}
	C(t)        &= C_0e^{-\lambda t}.		\label{prob_decay}
\end{align}
$\lambda$ is the degradation factor of $C(t)$ expressed as,
\begin{align}
	\lambda &= [S][E] = \frac{\ln2}{\hl}.   \label{degrad_fac}
\end{align}
$\hl$ corresponds to the half-life of the enzyme, which has a core role in controlling a constant amount of enzymes amongst different deployment scenarios. This is elaborated in detail in later sections. 

The mathematical expression for enzymatic reactions defines the probabilistic nature of degradation. It can be applied to our MCvD channel by using probability logic. If the function for arrival of molecules (hitting probability) at time $t$ to the receiver is $f_A(t)$, and the probability for degradation time $T$ being greater than arrival time $t$ is $P_B (T>t)$, the probability of messenger molecules hitting Rx before degradation becomes
\begin{align}
f_A(t)\cdot P_B(T>t)
\label{prob_degrade}
\end{align}
which is denoted by $h(t | \lambda)$ and equals to
\begin{align}
h(t | \lambda) = \frac{\rrn}{\dist+\rrn}\frac{\dist}{\sqrt{4\pi Dt^3}}e^{-\frac{\dist^2}{4Dt}-\lambda t}. \label{final_degrade}
\end{align}
The mathematical formula in \eqref{final_degrade} represents the hitting probability for a enzymatic MCvD channel with a point Tx and an absorbing sphere Rx. This will be used for analysis of scenarios with a point Tx, but does not directly correspond to cases for a sphere Tx. By indirectly using \eqref{final_degrade}, exponential decay can easily be implemented and simulated for the sphere Tx scenarios as well.

\section{Channel Environments} 
\subsection{Topology}
This paper considers two different topologies: point Tx to sphere Rx and sphere Tx to sphere Rx. For the Rx, a sphere shape is preferred to a point shape since better reception can be done with bigger shapes to a certain extent~\cite{Yilmaz20143Drx,Akka2015SIGREC}. For Tx, however, it is not yet clear which shape will be better for ISI mitigation. Therefore, the point and sphere Tx with identical enzyme area deployed around each of them are compared to see which is better for ISI mitigation.

\subsection{Channel Geometry and Parameters}
\begin{figure}[t]
\centering{\includegraphics[width=1\columnwidth,keepaspectratio]
{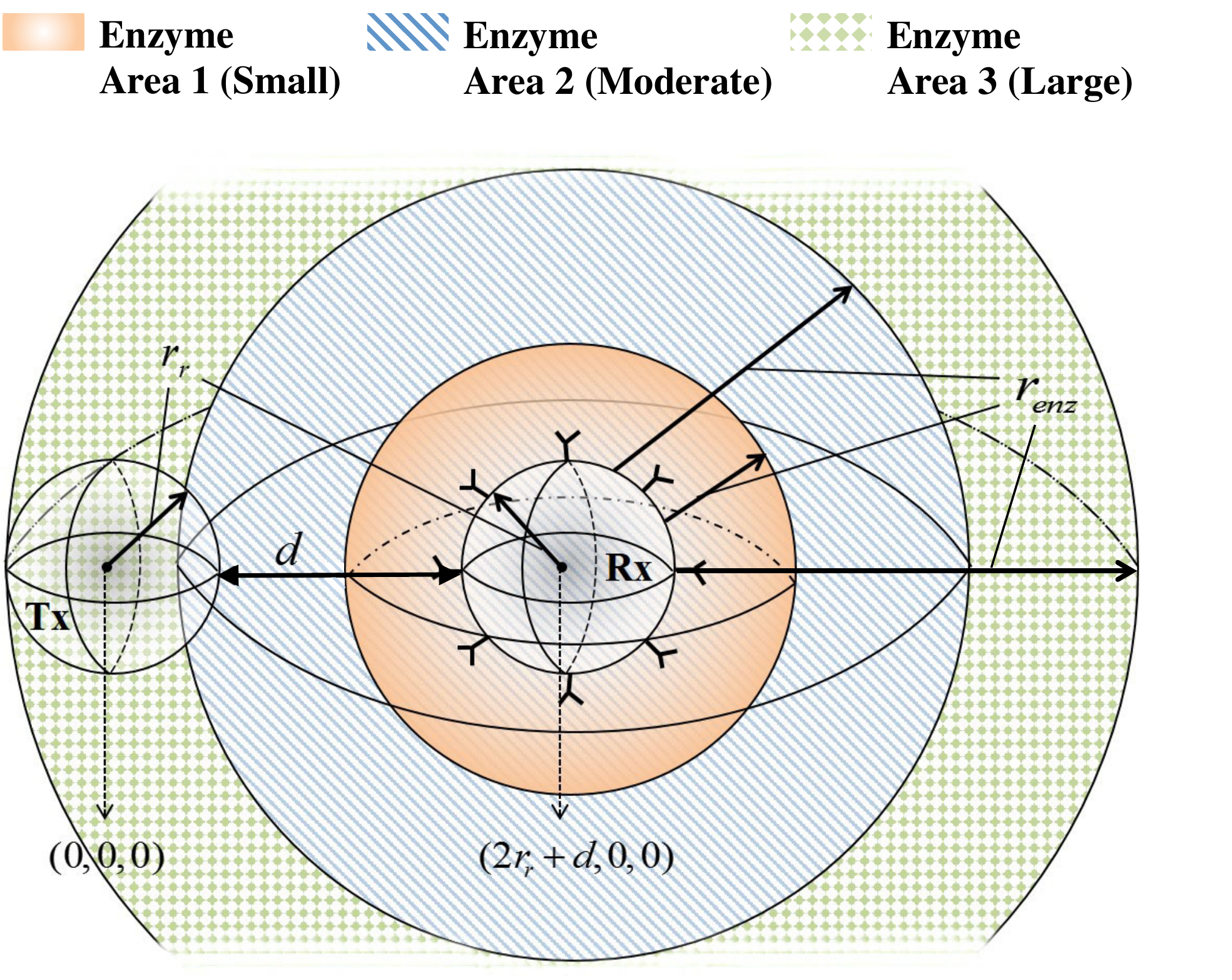}}
\caption{Detailed geometry and parameters of MCvD channel with enzymes deployed around Rx for three different sizes of enzyme area.}\label{fig:MCvDGeo}
\end{figure}

The specific geometry and important parameters of the MCvD channel are shown in Fig.~\ref{fig:MCvDGeo}. Figure~\ref{fig:MCvDGeo} shows enzymes deployed around the Rx with a sphere Tx for three different enzyme area cases. Other scenarios will have the same principle of geometry and system parameters with just different topology or type and size of the enzyme area. In Fig.~\ref{fig:MCvDGeo}, $\renz$ stands for the extended enzyme radius. The sphere Tx and Rx are both non-passive. The Tx reflects the messenger molecules that try to enter it by putting them back to their original positions and the Rx absorbs the messenger molecules that enter it by eliminating them from the channel after counting them. A point Tx will be passive in terms of interaction with the propagating molecules.

The enzyme area is an extending sphere shape being homocentric to the Rx or Tx, depending on the deployment structure. A limited amount of enzymes are only deployed within the enzyme area and the enzymes only affect the messenger molecules that are inside the designated enzyme area. Depending on the value of $\renz$, the enzyme area\ap\!s total volume will be decided. Note that the volume of the enzyme area is critical to implementing a constant number of limited enzymes in to different systems. In this study, the value of $\rrn$ is fixed and identical for both the Rx and Tx, but the $\renz$ and $\dist$ vary. The half-life of the enzymes is changed to see its\ap \! affect on the system.

\subsection{Limited Enzyme Implementation}
Since different channel scenarios with different enzyme area sizes are compared amongst each other, the amount of enzymes should always be kept constant for fair comparison. In order to keep the amount of enzymes identical for all of the scenarios, the volume of the enzyme area, $\vtot$, is used. Recall from \eqref{degrad_fac} that 
\begin{align}
	[S][E] = \ln2/\hl. \label{degrade}
\end{align}
If we fix the amount of enzymes to 1 and set $\ln2/[S]$ as constant $c$, $[E]$ is 
\begin{align}
	[E]=1/\vtot=c/\hl. \label{E}
\end{align} 
Therefore by multiplying a certain $\vtot$ value to $\hl$, a constant number of enzymes will be maintained amongst different enzyme areas and deployment scenarios. This special type of $\hl$ is the effective half-life explained in section~3.3.2.

\subsubsection{Total Enzyme Area}
The total enzyme area is needed to calculate the effective half-life. Since the total enzyme area, $\vtot$, should exclude any volumes of Tx or Rx that overlaps with the enzyme area, if $\vlp$ is the volume of the overlapping area, then 
\begin{align}
\label{eqn_vtot_general_form}
\vtot = \dfrac{4}{3}\pi\renz^3\,-\, \vlp \,.
\end{align}
For finding the value of $\vlp$, notice from Fig.~\ref{fig:MCvDGeo} that $\vlp$ changes depending on the $\renz$. For a small $\renz$, that is, $\renz \leq \dist$, $\vlp$ only contains the volume of a single Rx or Tx as in \textit{Enzyme Area 1} in Fig.~\ref{fig:MCvDGeo}. When $\renz$ increases and is within the range of $\dist+2\rrn > \renz > \dist$, $\vlp$ is the volume of a single Rx or Tx plus the lens-similar shape where the enzyme area and the Tx or Rx overlaps partially. This lens-similar shape is a sphere-to-sphere intersection and can be calculated accordingly~\cite{Kern1948sph2sph}. This second case corresponds to the case of \textit{Enzyme Area 2} in Fig.~\ref{fig:MCvDGeo}. The last case of $\vlp$ is when $\renz \geq \dist+ 2\rrn$. In this case $\vlp$ is the volume of both Tx and Rx since the enzyme area overlaps with both (\textit{Enzyme Area 3} in Fig.~\ref{fig:MCvDGeo}). Hence $\vlp$ is,
\begin{align}
 \vlp &= \begin{cases}
 			\dfrac{8}{3}\pi\rrn^3\,, \qquad \qquad \qquad \qquad \; \; \; \; \text{if } \renz \geq \dist + 2\rrn \\[9pt]
 			\dfrac{4}{3}\pi\rrn^3\,, \qquad \qquad \qquad \qquad \; \; \; \; \text{if } \renz  \leq \dist \\[9pt]
 			A + \dfrac{\rrn^2 - \renz^2}{4\dist_c}+ \dfrac{4}{3}\pi\rrn^3\,, \qquad   \text{otherwise}
 		\end{cases} \\
 	A &= \pi(\renz-\dist)^2\,\dfrac{\dist_{c}^2 + 2\dist_c\rrn -3\rrn^2 + 2\dist_c\Renz }{12\dist_{c}}\,, \nonumber \\
 	\dist_c & = 2\rrn + \dist \,, \quad \quad \Renz   = \rrn + \renz \,. \nonumber
\end{align}

\begin{figure*}[t]
\centering{\includegraphics[width=\textwidth,height=\textheight/3,keepaspectratio]{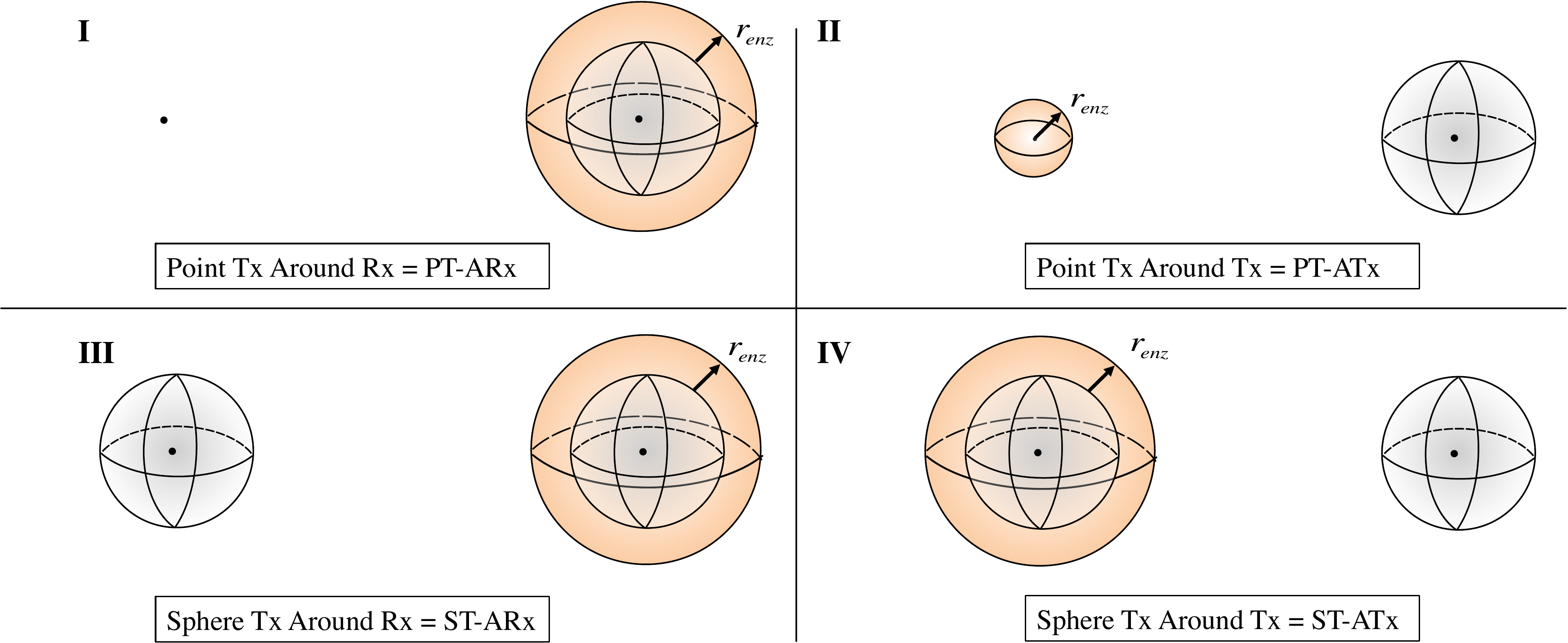}}
\caption{Diagram and notations of the four different enzyme deployment scenarios.}  \label{fig:MCvD_SCE}
\end{figure*}

Now we can calculate $\vtot$ in \eqref{eqn_vtot_general_form} which is utilized to evaluate the effective half-life for controlling a constant and limited number of enzymes.

\subsubsection{Effective Half-Life} 
To utilize \eqref{E}, where $[E]$ is inversely proportional to $\vtot$, a $\vtot$ must be multiplied to a reference $\hl$. Since our system\ap\!s $\vtot$ changes depending on the scenario type and $\renz$, we calculate a standard $\vtot$, denoted as ${V_{\text{totenz}, \onemicro}}$, and divide the current $\vtot$ with ${V_{\text{totenz}, \onemicro}}$ and multiply the result to the original half-life. The result is the effective half-life as in \eqref{effective_hl}.

Assume a system where $\renz = r_i$, half-life is $\Lambda^{r_i}_{1/2}$ with $V_{\text{totenz}, r_i}$. Then for two different cases of $\renz$ the half-life can be evaluated as, 
\begin{align}
\hl^{r_2} = \hl^{r_1} \frac{V_{\text{totenz}, r_2}}{V_{\text{totenz}, r_1}}. \label{hl_eq}
\end{align}

Now we define references $\hl$ and $\vtot$ as $\hle$ and $V_{\text{totenz}, \onemicro}$, which is the standard half-life and standard total enzyme area when $\renz = \onemicro$. This way for any different $\vtot$ we can calculate the effective half-life as
\begin{align}
\hl^{\renz} = \hle \frac{V_{\text{totenz}, \renz}}{V_{\text{totenz}, \onemicro}}. \label{effective_hl}
\end{align}

The effective half-life will accordingly change each time the scenario or $\renz$ changes. Note that enlarging the enzyme area reduces the degradation effect of enzymes due to the lowered enzyme concentration. On the other hand, enlarging the enzyme area also increases the probability of the diffusing molecules entering the enzyme area. Hence, there is a tradeoff between effectiveness of the enzymes and the probability of the molecules entering the enzyme area. This tradeoff suggests we should focus on finding the optimizing deployment scenario and $\renz$.

Substituting $\hl^{\renz}$ into \eqref{prob_decay}, we get the final probability of not decaying for each $\Delta t$ step for one messenger molecule inside the specified enzyme area as \eqref{prob_decay_renz}.  Now we have formulated a degrading function for the limited number of enzymes case in a specified enzyme area. 
\begin{align}
\mathbf{P} (\mbox{no degradation}\,| \hl^{\renz}) \!=\! e^{-\frac{ln(2)}{\hl^{\renz}}\Delta t} \!=\! \frac{1}{2^{\Delta t / \hl^{\renz}}} \label{prob_decay_renz}
\end{align}

\subsection{Enzyme Deployment Scenarios}
There are mainly four different enzyme deployment scenarios analyzed in this paper: Point Tx Around Rx, Point Tx Around Tx, Sphere Tx Around Rx, and Sphere Tx Around Tx as depicted in Fig.~\ref{fig:MCvD_SCE}. For the rest of this paper, these are named as PT-ARx, PT-ATx, ST-ARx, and ST-ATx, respectively. These types of scenarios are compared amongst each other while having identical $\renz, \, \hle, \, \ts$ (symbol period), and $\dist$ to make the channel environments identical except for the deployment type. Once the deployment with the best ISI mitigating performance is founded, the optimum $\renz$ value for different $\hle, \, \ts, \,$ and $\dist$ will be analyzed.

\section{Simulation System}
In our simulation system, for each time frame $\Delta t$,  every molecule emitted by the Tx moves by diffusion dynamics governed by the Gaussian distribution at each dimension, as follows
\begin{align}
\begin{split}
\Delta\vec{r} &= (\Delta x, \,\Delta y, \,\Delta z)  \\
\Delta x &\sim \mathcal{N}(0, \, 2D\Delta t) \\
\Delta y &\sim \mathcal{N}(0, \, 2D\Delta t) \\
\Delta z &\sim \mathcal{N}(0, \, 2D\Delta t)
\end{split}
\end{align}
where $\Delta\vec{r}$, $\Delta x$, $\Delta y$, and $\Delta z$ correspond to the displacement vectors and the displacements at $x$, $y$, and $z$ dimensions at a time frame of $\Delta t$ and $\mathcal{N}(\mu, \sigma^2)$ corresponds to the Gaussian distribution with mean $\mu$ and variance $\sigma^2$.

At each $\Delta t$ time step, each molecule is checked if it is inside the Tx node. If so, the molecule is put back to its\ap \,\,\!original position which is outside Tx. Each molecule is checked again to see if it is inside the Rx node. The ones inside the Rx are counted and eliminated~\cite{Birk2014SimStudy}, constituting the received signal. The last step for the simulation is to check for degradation of the remaining messenger molecules. For each molecule, the probability for not degrading \eqref{prob_decay_renz} is compared to a uniformly distributed random number for degradation check. This process is repeated until we reach $\tend$, the simulation end time.

\begin{table}[t]
\begin{center}
\caption{Values and ranges of the parameters used in the simulations.}
\renewcommand{\arraystretch}{1.2}
\label{tbl_system_parameters}
\begin{tabular}{p{4cm} l}
\hline
\bfseries{Parameter} 							& \bfseries{Value} \\ 
\hline 
 Diffusion Coefficient ($D$) 		& $\SI{100}{\square\micro\meter\per\second} $ \\ 
 Radius of the Rx/Tx ($\rrn$)   		& $\SI{5}{\micro\meter}$         			        \\
 Enzyme Radius ($\renz$)   		& $2 \sim \SI{26}{\micro\meter}$    \\ 
 Distance ($\dist$)				& $4, 6, 8, \SI{10}{\micro\meter}$\\
 Molecules Emitted for one $\ts$	& $5 \times 10^4$molecules\\ 
 Symbol Period ($\ts$)			& $0.1 \sim \SI{1.0}{\second}$\\
 Simulation End Time ($\tend$)		& $0.4, \SI{2.0}{\second}$\\
 Unit Half Life ($\hle$)			& $0.002 \sim \SI{0.008}{\second}$\\
 Simulation Step ($\Delta t$)	 	& $\num[retain-unity-mantissa = false]{1e-5} \, \si{\second} $\\
 Replications for Simulation		& 50\\
 \hline
\end{tabular} 
\end{center}
\end{table}
\renewcommand{\arraystretch}{1}

\section{Results and Analysis}
\subsection{Performance Metrics and Parameters}

For each simulation type, 50 replications are done. In our simulations different $\renz$ values are considered with fixed $\rrn$ for both Tx and Rx. For every different $\renz$, a different $\hl^{\renz}$ is calculated for maintaining a constant amount of limited enzymes and different $\mathbf{P} (\mbox{no degradation}\,|\, \hl^{\renz})$ will be applied to the system. 

For the evaluation of ISI, this study uses the interference-to-total-received molecules (ITR) metric. For a certain symbol period $\ts$, and simulation end time $\tend$, ITR is defined as:
\begin{align}
\text{ITR}(\ts, \tend) = \frac{F(\tend) - F(\ts)}{F(\tend)} 
\end{align}
where $F(\cdot)$ indicates the total number of molecules received until time $t$. The parameter indicates the portion of ISI molecules to the total number of received molecules. In our case, a smaller ITR indicates a better ISI mitigation. In Table~\ref{tbl_system_parameters}, we present the system parameters and their values or ranges that are used for the simulations and performance analysis. 

\subsection{Using Enzymes}
\begin{figure}[t]
\centering{\includegraphics[width=1\columnwidth,keepaspectratio]
{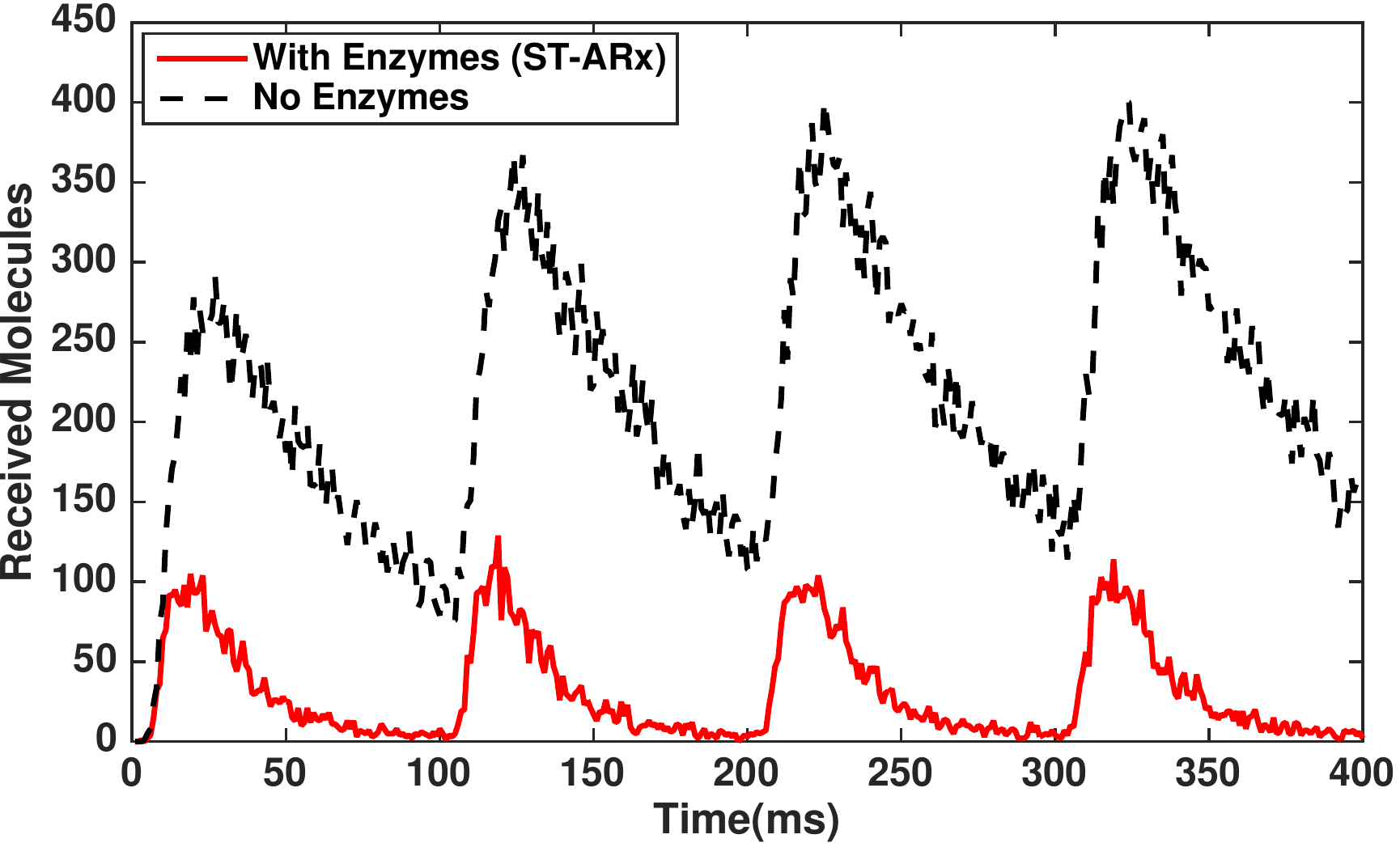}}
\caption{Received signals for ST-ARx system for four symbol periods when $\ts=\SI{0.1}{\second}$. \text{(${\dist=\SI{4}{\micro\meter}}$, ${\rrn=\SI{5}{\micro\meter}}$}, \text{${\renz=\SI{8}{\micro\meter}}$, ${\tend=\SI{0.4}{\second}}$, ${\hle=\SI{0.002}{\second}}$)}.}  \label{fig:SIG_NTS}
\end{figure}

The received signals for four symbol periods when enzymes are used and not used are shown in Fig.~\ref{fig:SIG_NTS}. In the received signal  for using enzymes, the ISI molecules do not accumulate so the height of the peak and tail of the signal is almost constant and small for all four symbol periods. On the contrary, when enzymes are not used, ISI molecules accumulate for each symbol period, causing the heights of the peak and tail of the signal to radically increase for each symbol period. This will more likely cause the receiver to erroneously decode the signal. Hence using enzymes prove to be more effective in ISI mitigation than not using them.

\begin{figure}[t]
\centering{\includegraphics[width=1.\columnwidth,keepaspectratio]
{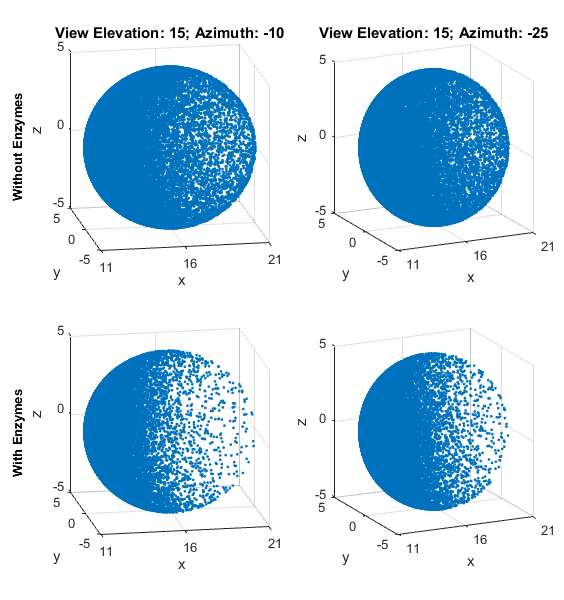}}
\caption{Molecule absorption locations at the receiver for \enquote{Without Enzymes} and \enquote{With Enzymes} \text{(${\dist=\SI{6}{\micro\meter}}$, ${\rrn=\SI{5}{\micro\meter}}$}, \text{${\renz=\SI{10}{\micro\meter}}$, ${\tend=\SI{2}{\second}}$, ${\hle=\SI{0.002}{\second}}$)}.}  \label{fig:antenna_pattern}
\end{figure}
To get more understanding of how enzymes affect the hitting probability, the point of hits for both of the cases, namely with and without enzymes is analyzed. Figure~\ref{fig:antenna_pattern} shows the hitting locations from different view points. Upper and lower rows correspond to the cases without and with enzymes, respectively. More molecules are hitting from the receiver\ap\!s back hemisphere for the without enzymes case compared to the enzyme added scenario. Molecules that are hitting from the back lobe travel longer distance than the other molecules which results in longer duration for reaching the receiver. Therefore, we can claim that the ISI is reduced when enzymes are utilized.

\subsection{Shape of Transmitter Node}
The topology of the enzymatic MCvD channel is analyzed to decide whether to use a sphere Tx or a point Tx. We compared PT-ATx and ST-ATx for determining which is better in terms of ITR. It is clearly supported by \eqref{hitp} that the hitting probability increases with increasing the receiver radius, so the Rx will remain as a sphere instead of a point. For the analysis we keep the $\rrn$ fixed and only focus on the Tx\ap\!s shape.

\begin{figure}[t]
\centering{\includegraphics[width=1\columnwidth,keepaspectratio]
{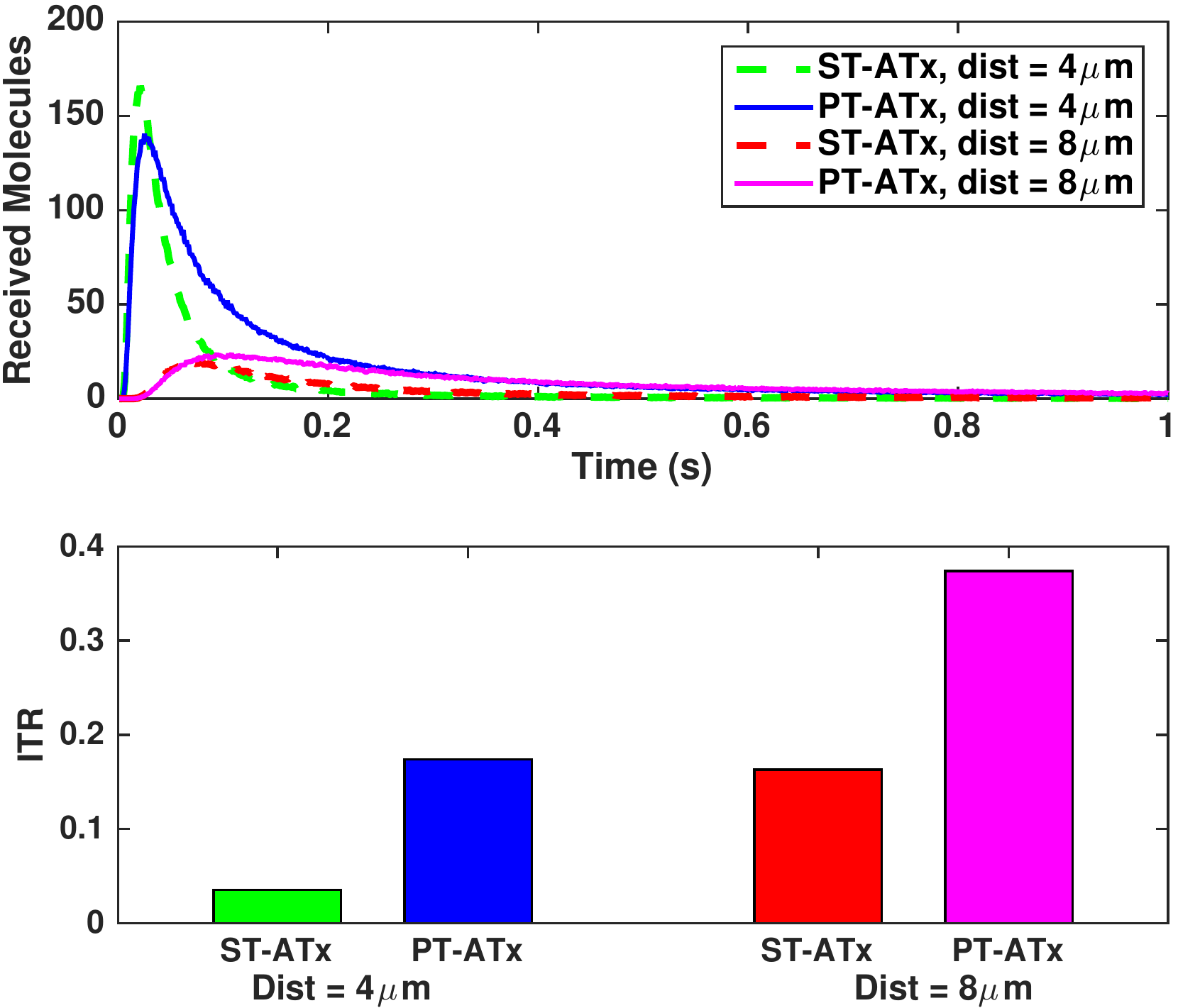}}
\caption{Received signals (Top) and ITR (Bottom) for comparing sphere and point source scenarios (${\rrn=\SI{5}{\micro\meter}}$, $\renz=\SI{2}{\micro\meter}$, ${\ts=\SI{0.5}{\second}}$, ${\tend=\SI{2.0}{\second}}$, ${\hle=\SI{0.002}{\second}}$).}  \label{fig:STx_PTx}
\end{figure}

Figure~\ref{fig:STx_PTx} shows how the received signal and ITR for ST-ATx and PT-ATx differs from each other for distance $\SI{4}{\micro\meter}$ and $\SI{8}{\micro\meter}$ when each scenarios\ap\,\!\! other system parameters were kept identical. Clearly, the signals for PT-ATx for both distances have a heavier tail than the ST-ATx signals. The ITR for ST-ATx is much lower than PT-ATx for both of the distances. Hence using a sphere Tx shows better ISI mitigation performance, and we will use a sphere Tx node for the rest of the analysis.

\subsection{Deployment Location}
After deciding that the sphere transmitter causes less ISI, the second decision parameter is about the enzyme deployment location. We analyze the performance of the following deployment locations, namely ST-ARx, ST-ATx, and \enquote{everywhere (randomly spread)}\footnote{Note that we need to use a limited enzyme area not to have zero enzyme concentration. Hence a considerably big area is used to refer to the case of, spread randomly \enquote{everywhere}. For instance we consider the enzyme radius four times the longest Tx-Rx  distance.}. With a limited amount of enzymes, whether enzymes should be densely deployed in a specific structure like ST-ARx and ST-ATX or just randomly spread around the entire channel like \enquote{everywhere} is unclear. In either cases we use the same amount of limited enzymes. If randomly spreading the enzymes yields better ISI mitigation than the other densely deploying scenarios, then pre-deciding a specific structure and area for the enzyme deployment will be unnecessary.

\begin{figure}[t]
\centering{\includegraphics[width=1\columnwidth,keepaspectratio]
{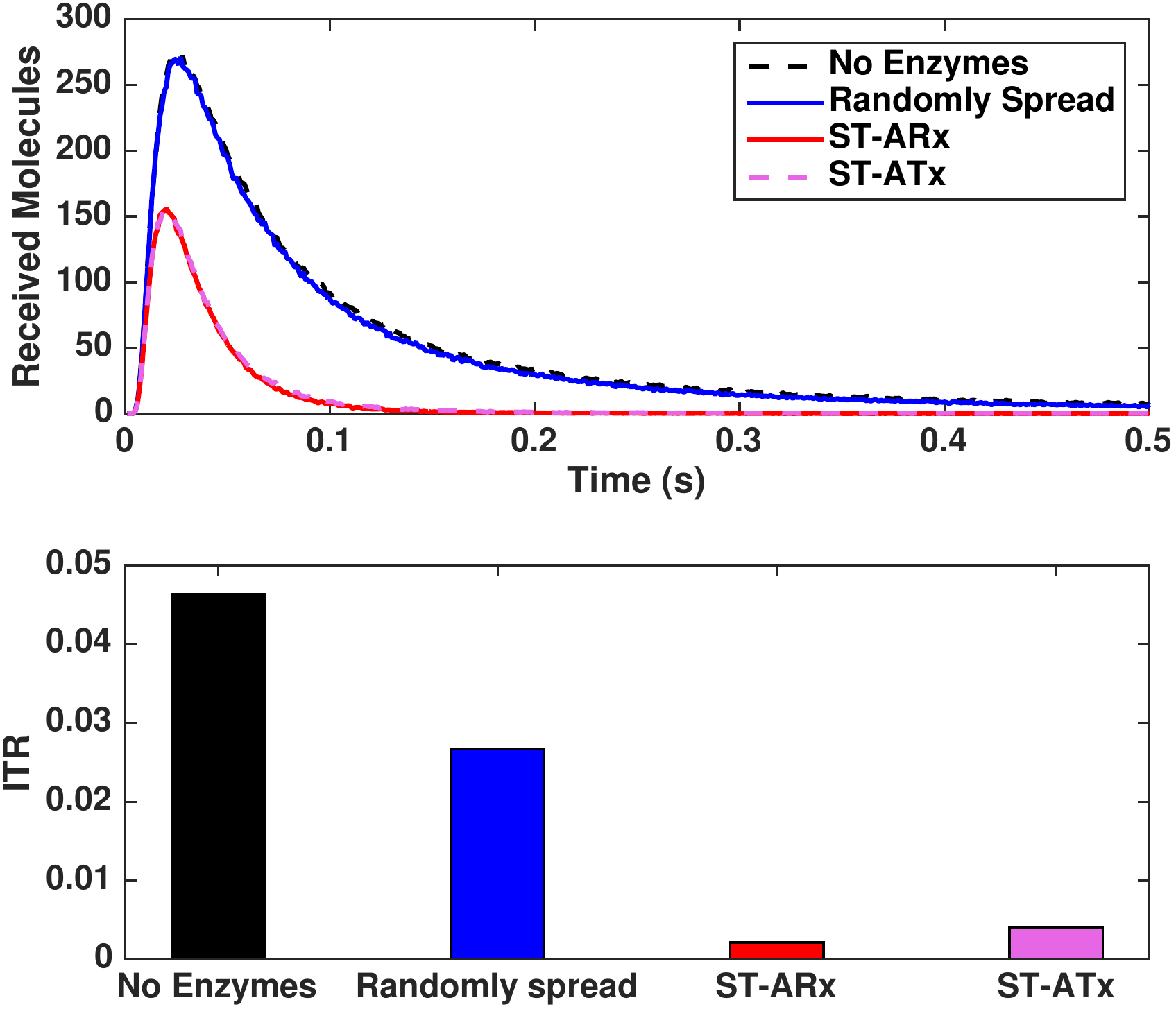}}
\caption{Received signals (Top) and ITR (Bottom) for comparing deployment schemes (${\dist=\SI{4}{\micro\meter}}$, ${\rrn=\SI{5}{\micro\meter}}$, ${\renz=\SI{6}{\micro\meter}}$, ${\ts=\SI{1.0}{\second}}$, ${\tend=\SI{2.0}{\second}}$, ${\hle=\SI{0.002}{\second}}$).}  \label{fig:EV_ENZ}
\end{figure}
Results in Fig.~\ref{fig:EV_ENZ} show that allocating enzymes in a specific structure, ST-ARx and ST-ATx, has lower ITR than just randomly spreading them everywhere. Spreading a certain amount of enzymes randomly around the channel has a received signal almost identical to that of using no enzymes. This implies that when the enzymes are spread out randomly throughout the channel, the amount of enzymes is so low compared to the entire volume of the channel that the channel is almost identical to that of \enquote{No Enzymes}. Hence, when a limited amount of enzymes is used, allocating them in a specific structure has lower ITR than randomly allocating them. ST-ARx and ST-ATx exhibit similar performance with the given parameters. The specific allocation structure that has better ITR property between ST-ARx and ST-ATx is analyzed more throughly in the next section.

\subsection{Deployment Structure: Around Rx/Tx}
\begin{figure}[t]
\centering{\includegraphics[width=1\columnwidth,keepaspectratio]
{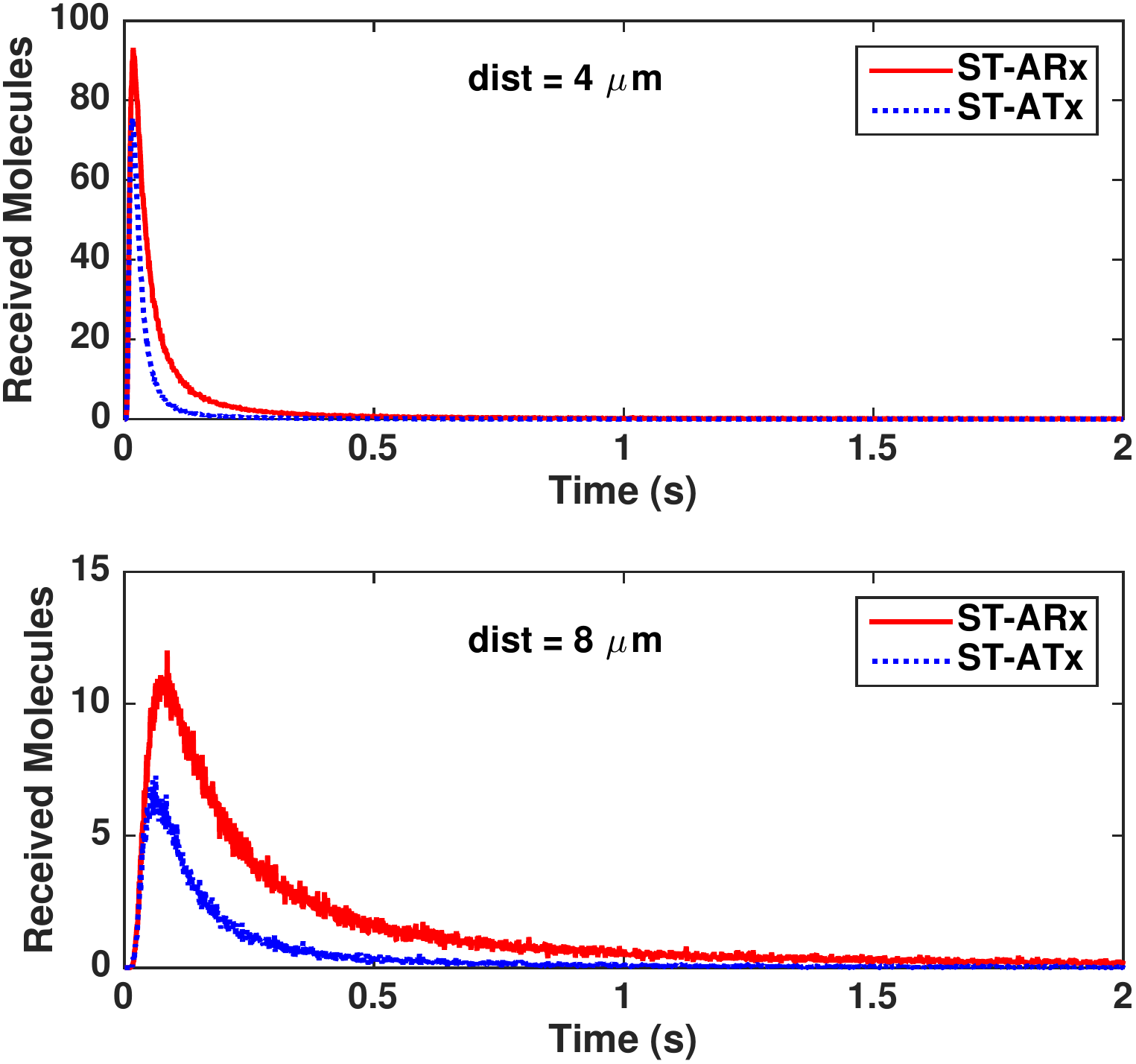}}
\caption{Received signals of ST-ARx, and ST-ATx scenarios for $\dist \,= \SI{4}{\micro\meter}$ (Top) and $\SI{8}{\micro\meter}$ (Bottom) (${\rrn=\SI{5}{\micro\meter}}$, $\renz=\SI{2}{\micro\meter}$, ${\tend=\SI{2.0}{\second}}$, ${\hle=\SI{0.002}{\second}}$).}  \label{fig:RXTX_SIG}
\end{figure}

In general, ISI molecules are considered to accumulate closer to the Rx than the Tx after propagating some distance. ST-ARx may therefore be assumed to give better ISI mitigation. To evaluate this assumption, ST-ARx and ST-ATx are compared for two distances (4 and $\SI{8}{\micro\meter}$) with different $\ts$ and $\renz$. Figure~\ref{fig:RXTX_SIG} shows the received signals for each of the scenarios. The difference is clear between the signals in terms of signal peak and the heaviness of the signal tail. For both distances, ST-ATx has the signal with lower peak and shorter, less-heavy tail than that of ST-ARx. 

\begin{figure}[t]
\centering{\includegraphics[width=1\columnwidth,keepaspectratio]
{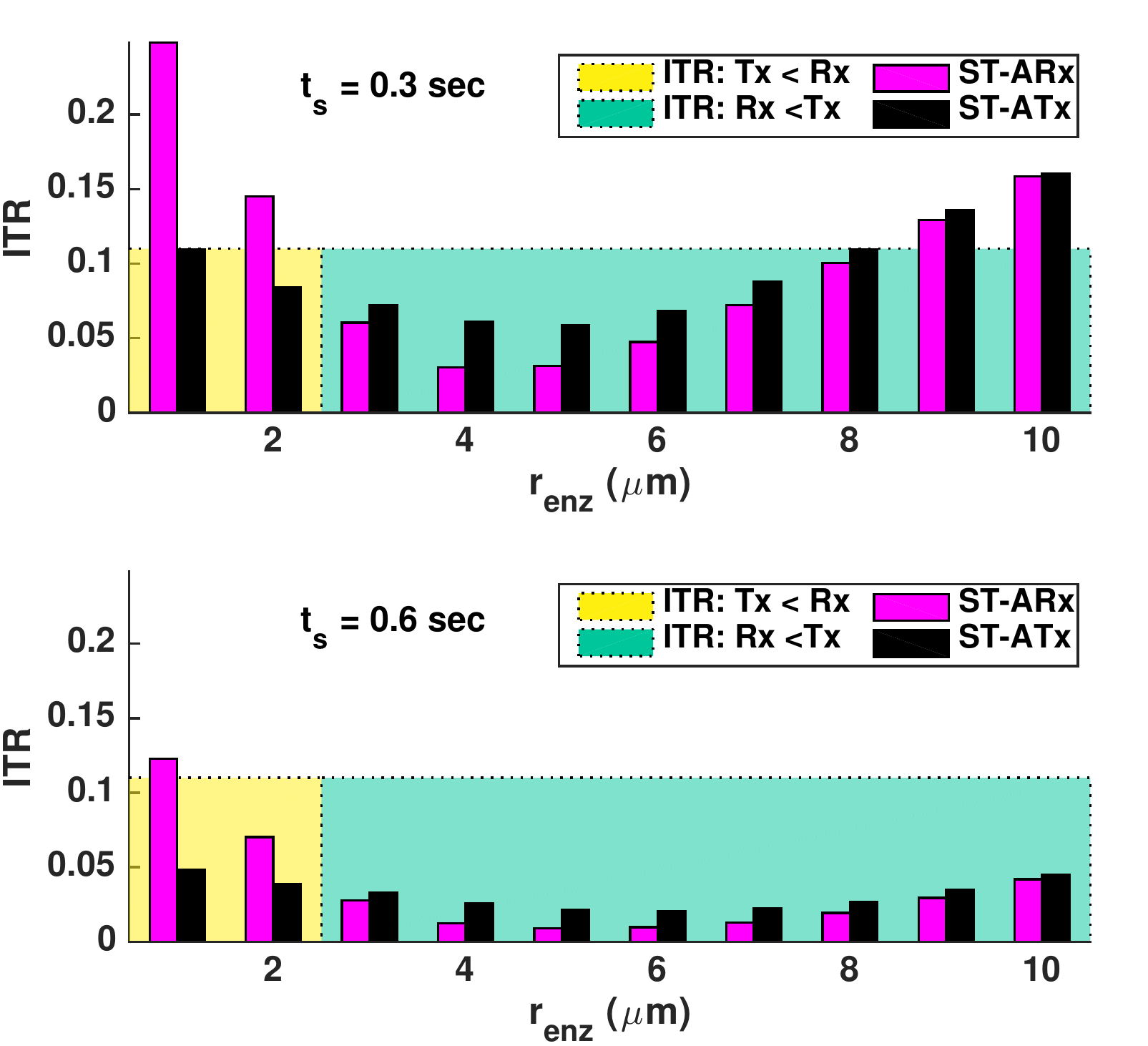}}
\caption{ITR of ST-ARx, and ST-ATx scenarios for $\ts\,=\,0.3\,{\text{sec}}\;$ (Top) and $0.6\,{\text{sec}}$ (Bottom) and different $\renz$ (${\dist=\SI{4}{\micro\meter},\; \SI{8}{\micro\meter}}$, ${\rrn=\SI{5}{\micro\meter}}$, ${\tend=\SI{2.0}{\second}}$, ${\hle=\SI{0.002}{\second}}$).}  \label{fig:RXTX_ITR}
\end{figure}

More analysis is done with more varied system parameters in the ITR graph in Fig.~\ref{fig:RXTX_ITR}. Figure~\ref{fig:RXTX_ITR} shows ITR for ST-ARx and ST-ATX with different $\renz$ for $\ts=$ 0.3 sec and 0.6 sec. The ITR graphs show similar trends for both $\ts$ values. Until the $\renz$ reaches a certain value, $\SI{4}{\micro\meter}$ in this case, ST-ATx has lower ITR than ST-ARx. Once that value is exceeded, ST-ARx starts to have lower ITR than ST-ATx and reaches the lowest ITR value. Once $\renz$ gets large enough, however, the ITR of ST-ARx and ST-ATx are almost identical as both channels become similar to the channel in which enzymes are randomly spread everywhere.

When the enzyme area is tight, ST-ATx is better in ISI mitigation than ST-ARx. Hence, with the selected parameters, if the enzyme deployment constraints do not allow $\renz$ to be greater than $\SI{4}{\micro\metre}$, then deploying the enzymes around Tx should be selected. When, however, the enzyme area gets large to a certain value, around Rx is preferable. The two scenarios get nearly identical ITR when the enzyme area gets very large. The lowest ITR occurs for ST-ARx. Therefore when optimum ITR mitigation is necessary regardless of the enzyme area size, ST-ARx should be used. 

\begin{figure}[t]
\centering{\includegraphics[width=1\columnwidth,keepaspectratio]
{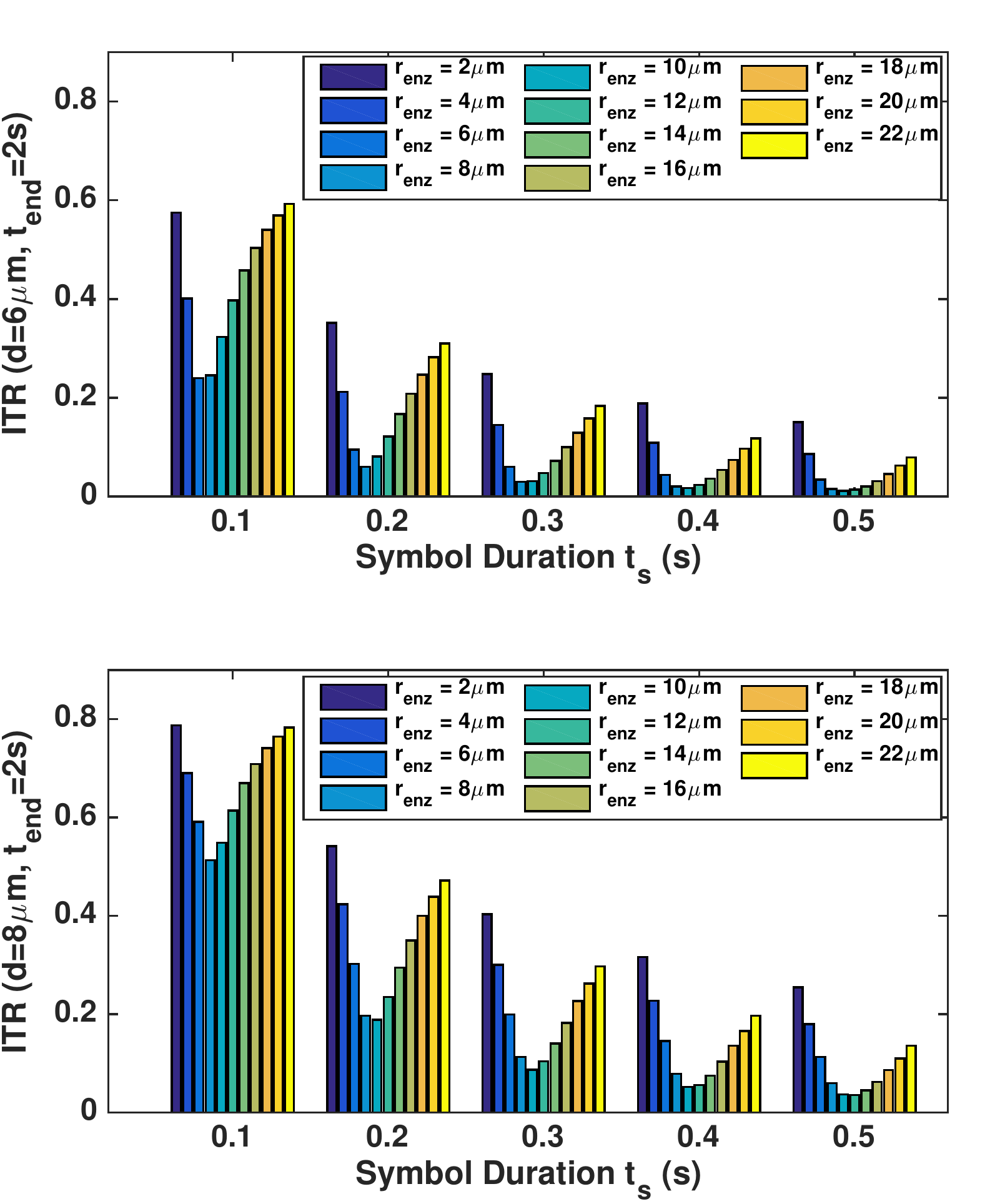}}
\caption{ITR of ST-ARx with varying $\renz$ and $\ts$ for ${\dist=\SI{6}{\micro\meter}}$ (Top) and $\SI{8}{\micro\meter}$ (Bottom) (${\rrn=\SI{5}{\micro\meter}}$, $\tend=\SI{2.0}{\second}$, ${\hle=\SI{0.002}{\second}}$).}  \label{fig:RX_OPTRENZ}
\end{figure}

The size of the enzyme area (i.e., $\renz$) that maximizes ISI mitigation for the ST-ARx scenario is also analyzed. Figure~\ref{fig:RX_OPTRENZ} shows the graph of the ITR for varying $\renz$ and $\ts$ for $\dist$ = $\SI{6}{\micro\meter}$ and $\SI{8}{\micro\meter}$. Clearly there is an optimum $\renz$, \, namely $\optr$, where lowest ITR occurs for each $\dist$ and $\ts$. Here, $\optr$ is defined as the $\renz$ when lowest ITR occurs for the specific channel. In cases of Fig.~\ref{fig:RX_OPTRENZ}, $\optr$ ranges from 6 - \SI{12}{\micro\meter} depending on the distance and $\ts$. How $\optr$ is influenced by the channel parameters $\dist$, $\ts$, and $\hle$ is elaborated in the next section.

\subsection{Relation of $\optr$ to Channel Parameters }

\begin{figure}[t]
\centering{\includegraphics[width=1\columnwidth,keepaspectratio]
{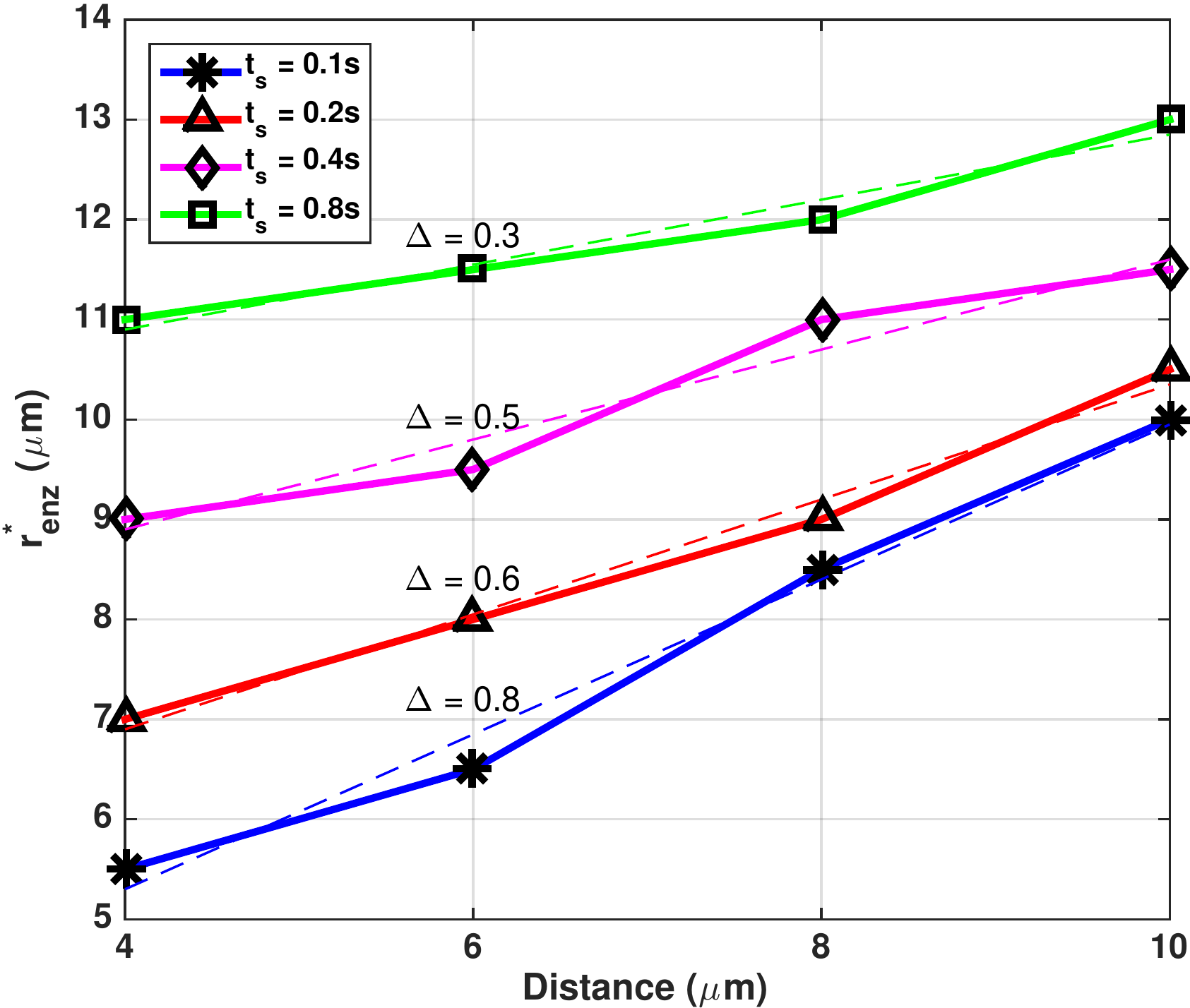}}
\caption{$\optr$ depending on the distance for different $\ts$ for ST-ARx scenario with fitting lines (${\rrn=\SI{5}{\micro\meter}}$,  ${\tend=\SI{2.0}{\second}}$, ${\hle=\SI{0.002}{\second}}$).}  \label{fig:OPTRENZ_DIST}
\end{figure}

This section analyzes how $\optr$ is related to the channel\ap\!s distance, symbol period, and half-life. To specify in detail the relation between the distance and $\optr$, Fig.~\ref{fig:OPTRENZ_DIST} presents the varying $\optr$ depending on the increasing $\dist$ for different $\ts$. It is clear from the graph that there is a steady, upward trend relationship between the distance and the $\optr$ for all $\ts$. If the distance increases this will mean that the $\optr$ also is increased, implying an optimum ratio of distance to $\optr$ for maximized ISI mitigation. The slope of the fitting lines, $\Delta$, for each of the $\ts$ is also shown, suggesting that as $\ts$ increases geometrically by a multiplication of two, the increase of $\optr$ gets less steep. Therefore for an increasing distance, $\renz$ also must be increased for optimizing ISI mitigation but the symbol period, $\ts$, should also be taken into consideration regarding to how steeply $\optr$ changes.

\begin{figure}[t]
\centering{\includegraphics[width=1\columnwidth,keepaspectratio]
{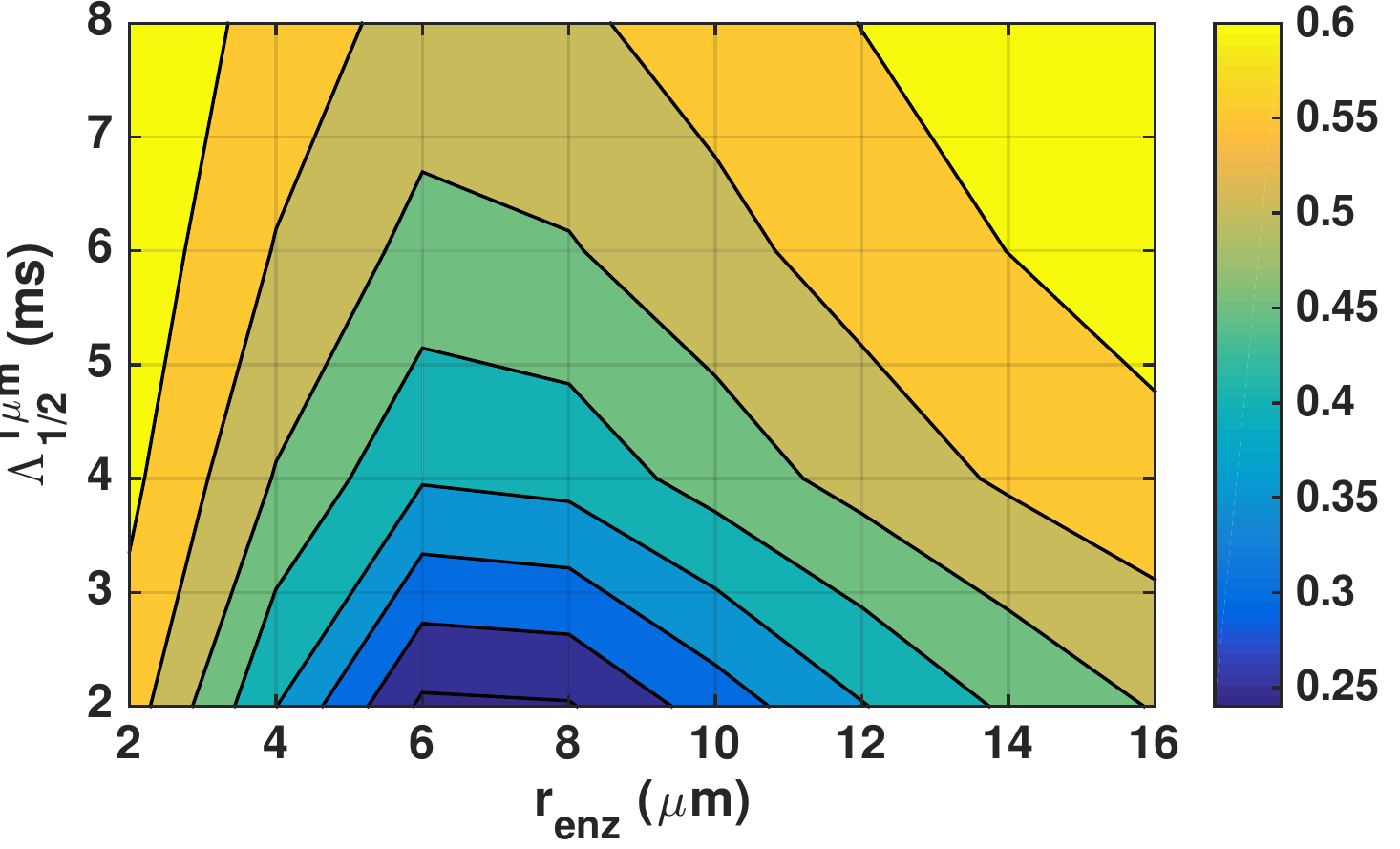}}
\caption{Heatmap of ITR with varying $\renz$ and $\hle$ for ST-ARx scenario ($\dist\,=\,\SI{6}{\micro\meter}$, ${\rrn=\SI{5}{\micro\meter}}$, ${\ts=\SI{0.1}{\second}}$, ${\tend=\SI{2.0}{\second}}$).}  \label{fig:RX_HALFLIFE_HEATM}
\end{figure}

$\optr$\ap\!s dependence on the unit half-life, $\hle$, is also evaluated. The half-life of an enzyme is defined as the time required for the enzyme\ap\!s target substrate concentration to fall to its\ap\,\! half value. Hence the lower the half-life the faster the enzyme degrades the substrates. Figure~\ref{fig:RX_HALFLIFE_HEATM} shows a heatmap of the ITR with $\renz$, $\hle$ as the $x$, $y$ axis respectively. Four different half-lives are considered in this study: 2, 4, 6, $\SI{8}{\milli\second}$.  Clearly, the lower the $\hle$ the lower the ITR since the degradation occurs faster. The $\optr$, however, does not change according to the $\hle$. For all four $\hle$ the $\optr$ is \SI{6}{\micro\meter} in this case. Therefore the $\hle$ affects only the rate of degradation but not the $\optr$ value.

\section{Conclusion}
This paper analyzed the different system structures and parameters that can maximize ISI mitigation with using a limited amount of enzymes. In terms of topology, when the same amount of enzymes were used a sphere Tx was shown to yield more ISI mitigation than a point Tx. For the enzyme deploying location, randomly deploying the enzymes everywhere created more ISI molecules than deploying them in a specific structure. As to which specific structure is more preferable, when the enzyme area is small to a certain extent, ST-ATx had less ISI. For a larger enzyme area, however, ST-ARx had less ISI and the lowest ISI occurred for the ST-ARx scenario. Once the enzyme area got very large, the two different scenarios yielded almost the identical results. 

For the case of ST-ARx there proved to be an optimum size of the enzyme area that appeared to maximize the ISI mitigation. This optimum enzyme area increased as the distance between the Rx and Tx increased, and the rate of increase lessened as the symbol period increased. The half-life, on the other hand, had no effect on the optimum enzyme area size, but a lower half-life meant less ISI.

Further research is possible on deriving the mathematical interpretations and expressions for the limited enzyme implementation with optimized system parameters. Moreover, the research can be applied to molecular MIMO systems~\cite{Lee2015MIMO, Bon2016MIMO} where the limited enzymes around Rx or Tx can be used as a methodology for mitigating inter-link interference.

\ack  \noindent This research was supported by the MSIP (Ministry of Science, ICT and Future Planning), Korea, under the ``IT Consilience Creative Program" (IITP-2015-R0346-15-1008) supervised by the IITP (Institute for Information \& Communications Technology Promotion) and by the Basic
Science Research Program (2014R1A1A1002186) funded by the MSIP, Korea,
through the National Research Foundation of Korea.


\bibliography{ettRefs}
\bibliographystyle{wileyj}

\end{document}